\documentclass[10pt,draft]{dis03}
\usepackage{epsf,amsmath}

\begin{document}

\newcommand{\pb}{\mbox{{\rm ~pb}}}
\newcommand{\Rp}{\mbox{$\not \hspace{-0.15cm} R_p$}}
\def\GeV{\hbox{$\;\hbox{\rm GeV}$}}

\title{SEARCHES FOR R-PARITY VIOLATING SUPERSYMMETRY AT HERA 
{\thanks{Talk given at the $XI^{th}$ International Workshop 
on Deep Inelastic Scattering (DIS'03),
23-27 April, St. Petersburg, Russia }} 
}

\author{E.~PEREZ, on behalf of the H1 Collaboration \\
CE-Saclay, DSM/DAPNIA/Spp, France } 

\maketitle

\begin{abstract}
\noindent Searches for R-parity violating supersymmetry performed
at the H1 experiment are presented. 
Emphasis is put on searches for squarks, which may be resonantly 
produced at the HERA $ep$ collider in supersymmetry where R-parity
is violated.
The preliminary results presented here were obtained using
data collected at a centre-of-mass energy of 320 GeV and corresponding to 
an integrated luminosity of 63 pb$^{-1}$.
\end{abstract}

\section{Introduction}

Searches for minimal supersymmetry (SUSY) at HERA with conservation of the so-called
R-parity ($R_p$) have been performed by both the H1 and ZEUS 
experiments~\cite{H1ZEUSMSSM}
at the HERA $ep$ collider, looking for the production of 
a selectron-squark pair. HERA's sensitivity
in such supersymmetric frameworks is however limited, when taking into
account the mass range allowed
by other experiments.

On the other hand, HERA is very well suited to search for squarks 
which possess R-parity violating ($\Rp$) interactions.
The resonant production of such squarks is adressed here, using H1 $e^+ p$ data
collected in 1999-2000 and corresponding to an integrated luminosity of 
$\sim 63 \pb^{-1}$.
Earlier squark searches performed using a lower statistics
sample of H1 $e^+ p$ data collected at $\sqrt{s} = 300 \GeV$
are published in~\cite{H1SUSYPAPER}.

\section{Phenomenology of $\Rp$ Supersymmetry at HERA}


The most general SUSY theory which preserves the gauge invariance
of the Standard Model (SM) allows for Yukawa couplings between
two known SM fermions and a squark or a slepton.
Such couplings do not conserve the R-parity, defined as
$R_p = (-1)^{3B+L+2S}$ where $S$ denotes the spin, $B$ the baryon number and
$L$ the lepton number of the particles.
The $\Rp$ interactions which are the most relevant for HERA~\cite{RPVIOLATION}
allow
the resonant production of $\tilde{u}_L^j$ and $\tilde{d}^{k}_R$,
respectively the SUSY partners of the left-handed $u^j$ and right-handed 
$d^k$ quarks, where $j$ and $k$ denote generation indices.
The relevant processes in $e^+ p$ collisions,
$e^+ d^k \rightarrow \tilde{u}_L^j$ and
$e^+ \bar{u}^j \rightarrow \tilde{d}^{k*}_R$, 
are mediated by the so-called $\lambda'_{1jk}$ coupling.
The sensitivity is thus highest for the 
$\tilde{u}_L^j$ production via $\lambda'_{1j1}$, due to the
most favorable parton density.
%


Although the strength of the coupling $\lambda'_{111}$ is severely constrained by
the non-observation of neutrinoless double-beta decay, no 
such strong bounds exist
on the other $\lambda'_{1j1}$ couplings, allowing for potential large 
cross-sections at HERA.
For example with $\lambda'_{1j1} = 0.3$ the
production cross-section of a 290~GeV $\tilde{u}_L^j$ squark
is $\sim 0.2$~pb for $\sqrt{s} = 320 \GeV$.
That is a factor of 
$\sim 10$ larger than the corresponding cross-section for $\sqrt{s} = 300 \GeV$.


When the produced squark undergoes an $\Rp$ decay the final state
consists of a lepton and a quark. 
Final states with a larger multiplicity arise from $R_p$ conserving decays of
the squark into a quark and a neutralino,
a chargino
or a gluino. These gauginos, including the lightest one assumed to be
the lightest supersymmetric particle (LSP), are not stable and decay either
directly via \Rp\ into a lepton and two jets (mainly relevant for the LSP), or 
into lighter gauginos. The various decay chains lead to final states with several
jets and one or several leptons.

\section{Analysis}


When the squark undergoes a \Rp\ decay into a quark and an 
electron,
the final state is similar to that of high $Q^2$ Neutral Current (NC)
Deep Inelastic Scattering (DIS), $Q^2$ denoting the square of the four-momentum
carried by the exchanged boson.
%
The characteristic angular distribution for the decay products of a scalar
particle is used to reduce the DIS background. 
The invariant mass in the centre-of-mass of the hard-subprocess is reconstructed
with a resolution of $3 - 6 \GeV$ for signal events.
The observed invariant mass distribution does not show any resonant
peak, and agrees well with the SM expectation up to the highest masses.

%

\begin{figure}[thb]
 \begin{center}
  \mbox{\epsfxsize=0.8\textwidth
       \epsffile{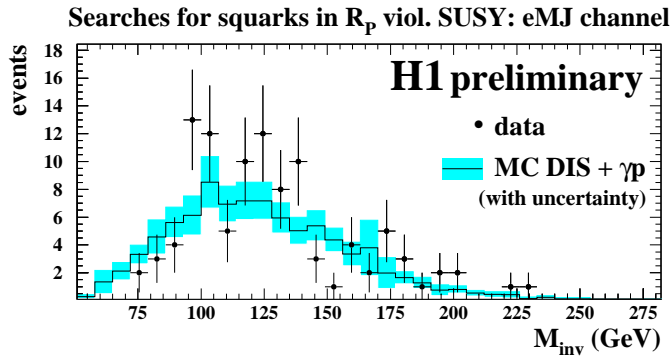}}
  \end{center}
 \caption[]{ \label{fig:dNdMeMJ}
   Distribution of the invariant mass of events with a positron and
   several jets in the final state.
   }
\end{figure}

Final states resulting from an $R_p$ conserving squark decay have been
classified into several final states, depending on the number and nature
of the final state lepton(s). 
As an example, Fig.~\ref{fig:dNdMeMJ} shows the observed
and expected invariant mass distributions for events with an $e^+$ and several jets
in the final state. Such final states might arise from the $\tilde{u}_L^j$ decay
into a neutralino, followed by the
$\Rp$ decay of the latter into a positron and two jets.
A good agreement is observed with the SM prediction. 
No striking event 
$e^+ p \rightarrow e^- + {\rm{ jets}} + X$ showing an explicit violation
of the lepton number (characteristic of \Rp\ induced by 
a $\lambda'$ coupling) has been observed.
The signal has also been looked for in channels with a neutrino and several jets,
and with several leptons ($ee$, $e \mu$, $e \nu$, $\mu \nu$) and several jets.
No discrepancy with the SM prediction has been observed.
%

\section{Constraints on SUSY models}

Upper bounds on the production cross-section of a $\tilde{u}_L^j$
are derived, combining all analyzed channels.
In SUSY models\footnote{
    More details about the models considered here can be found
    in~\cite{H1SUSYPAPER} and references therein.}
inspired from the Minimal Supersymmetric Standard
Model (MSSM), the channels considered in this analysis cover
$\sim 100 \%$ of the squark decay modes.
These bounds are translated into constraints on the parameters
of SUSY models.

\begin{figure}[thb]
 \begin{center}
  \begin{tabular}{cc}
  \mbox{\epsfxsize=0.45\textwidth
       \epsffile{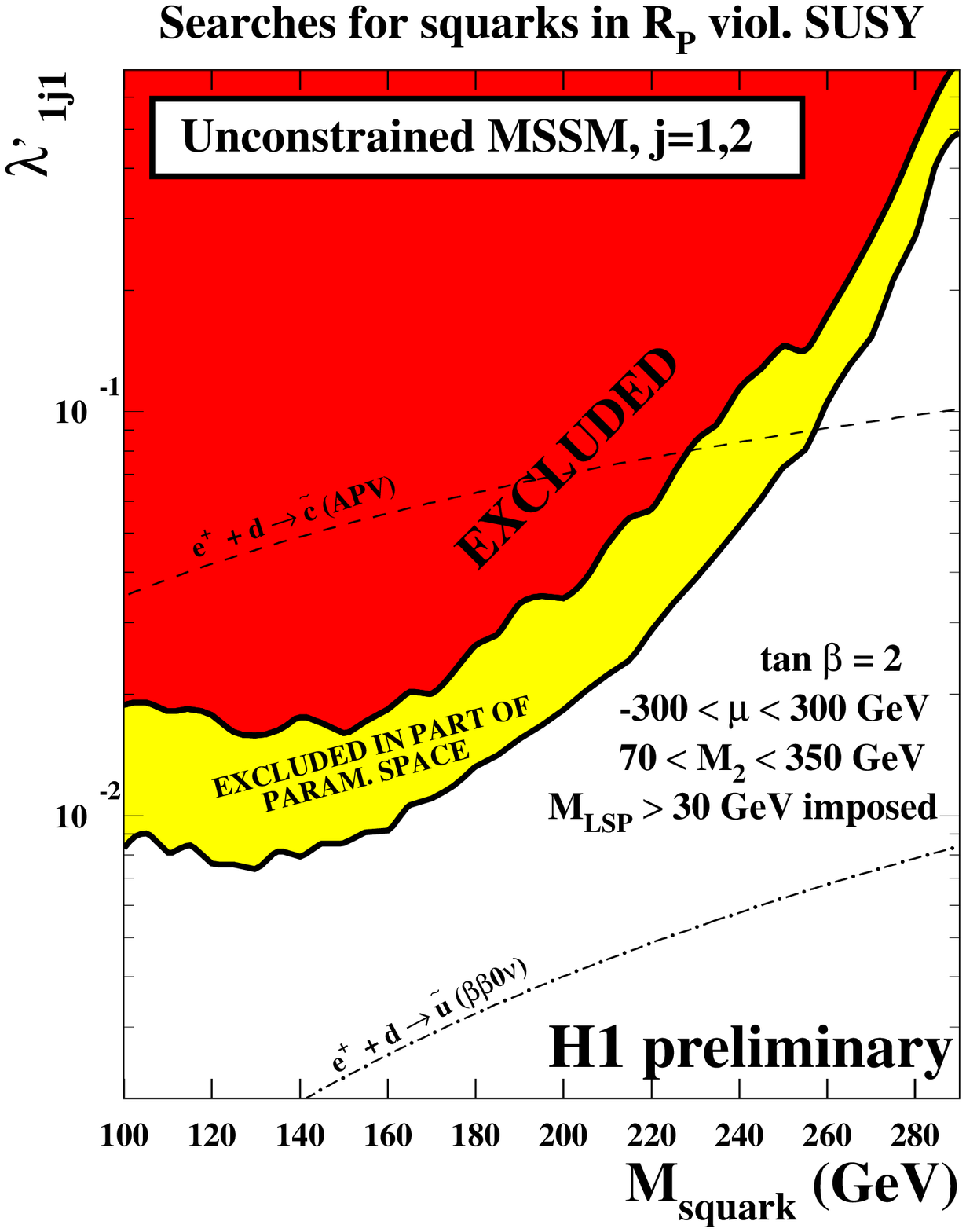}}
  &
  \mbox{\epsfxsize=0.45\textwidth
       \epsffile{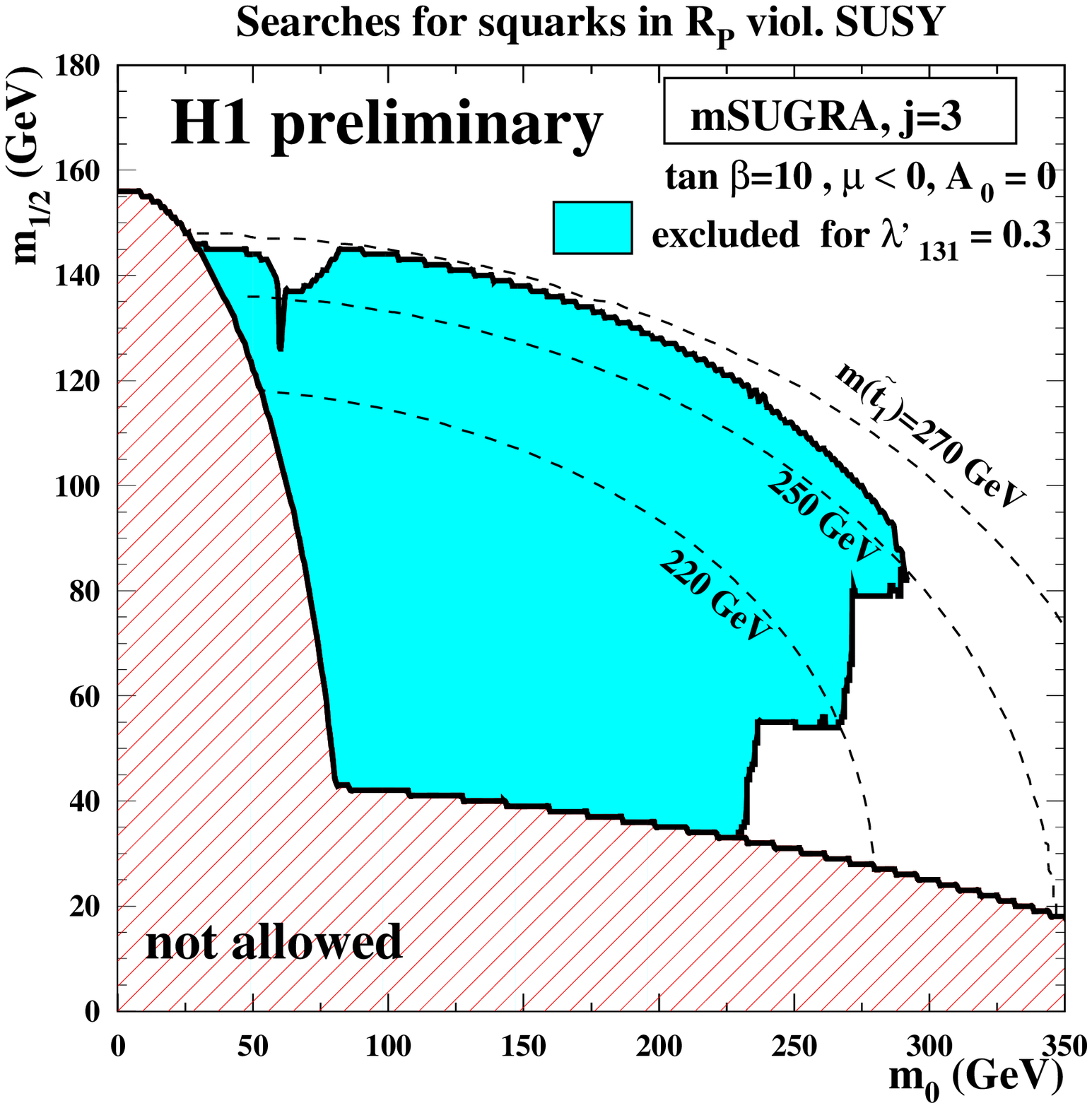}}
  \end{tabular}
  \end{center}
 \caption[]{ \label{fig:MSSM}
   (left) Upper limits for the coupling $\lambda'_{1j1}$ ($j=1,2$) as a function
   of the squark mass, in the ``phenomenological'' MSSM;
   (right) Constraints obtained in the mSUGRA model, assuming
   $\lambda'_{131} = 0.3$.
   }
\end{figure}

Fig.~\ref{fig:MSSM} (left) shows upper limits on the Yukawa coupling
$\lambda'_{1j1}$ for $j=1,2$ as a function of the $\tilde{u}^j_L$ mass.
These are obtained in a ``phenomenological" MSSM, where the gaugino
masses are related to each other while the sfermion masses are free.
A scan of the parameter space is performed, which shows that the obtained
limits do not depend strongly on the model parameters.
For a coupling of the electromagnetic strength ($\lambda' = 0.3$)
squark masses up to $\sim 280 \GeV$ are excluded.
Limits obtained on the coupling $\lambda'_{121}$ extend beyond the bounds
from low energy experiments. That is also the case for the coupling
$\lambda'_{131}$ (not shown), which could allow for resonant stop
production at HERA.

Example constraints obtained in the framework of the minimal Supergravity 
(mSUGRA) model are shown in Fig.~\ref{fig:MSSM} (right).
Here, a common mass $m_0$ ($m_{1/2}$) is assumed for the 
scalars (gauginos) at the Grand Unification scale.
Fig.~\ref{fig:MSSM} (right) shows the excluded domain in the 
($m_{1/2}$, $m_0$) plane which is obtained assuming a coupling
$\lambda'_{131} = 0.3$. Under this hypothesis stop masses up to
$\sim 270 \GeV$ can be ruled out. 
%
%
The future sensitivity of the Tevatron experiments on light stop
squarks might be around 200-250 GeV, depending on the main decay
modes of the stop.
A reasonably large coupling $\lambda'_{131}$ would thus provide
an interesting discovery potential for the stop at HERA, with
the much larger integrated luminosity expected within the next years.

\section{Other interesting topologies}

In the results presented above solely fermionic decays
of squarks were considered. 
In some SUSY models~\cite{KON} where e.g. the sbottom is light,
a resonantly produced stop might decay into a sbottom and a $W$
boson, with cross-sections of $\sim 1 \pb$ being possible. 
This would be a source of events with a hard lepton,
a hard jet and a large amount of missing transverse energy. 
A slight excess of such events is actually observed in the
H1 experiment~\cite{H1ISOLEP}.

Other interesting events with several electrons in the final state
have also been reported by H1~\cite{MULTILEP}. An interpretation of
these events involving the \Rp\ resonant production of a sneutrino
($ee \rightarrow \tilde{\nu}_{\mu, \tau} \rightarrow ee$) via
$e \gamma$ collisions 
is unlikely taking into account LEP bounds on such a process.

\section{Conclusions and Prospects}

Stringent bounds on $\tilde{u}^j_L$ squarks with R-parity 
violating interactions
have been set by the H1 experiment. The analysis 
of $e^- p$ data taken in 98-99 will bring complementary constraints
on $\tilde{d}^k_R$ squarks. 
The seven to ten-fold increase of integrated luminosity expected by 2006
could offer a large discovery potential for stop squarks provided
that the relevant \Rp\ coupling is not too small.

\end{document}